\documentclass[aps,pr,showpacs,floats,11pt]{revtex4}
\usepackage{graphicx}
\newcommand{\beq}{\begin{equation}}
\newcommand{\eeq}{\end{equation}}
\newcommand{\beqa}{\begin{eqnarray}}
\newcommand{\eeqa}{\end{eqnarray}}

\newcommand{\z}{\zeta}

\newcommand{\f}{\phi}

\renewcommand{\d}{\delta}
\renewcommand{\r}{\rho}

\renewcommand{\r}{\rho}
\renewcommand{\b}{\beta}
\renewcommand{\l}{\lambda}
\renewcommand{\inf}{\infty}
\renewcommand{\>}{\rangle}
\newcommand{\<}{\langle}

\begin{document}
\title{Mean Field and the Single Homopolymer}
\author{S. Pasquali$^1$, J.K. Percus$^2$}
\affiliation{$^1$Laboratoire de Physico-Chime Th\'eorique, UMR Gulliver CNRS-ESPCI 7083, 10
  rue Vauquelin, 75231 Paris Cedex 05, France.\\ $^2$Courant Institute and
  Physics Department NYU, 251 Mercer St. New York, NY 10012}
\date{\today}
\begin{abstract}
We develop a statistical model for a confined chain molecule based on a monomer grand canonical ensemble.
The molecule is subject to an external chemical potential, a backbone interaction, and an attractive
interaction between all monomers.
Using a Gaussian variable formalism and a mean field approximation, we analytically derive a 
minimum principle from which we can obtain relevant physical quantities, such as the monomer density,
and we explore the limit in which the chain is subject to a tight confinement.
Through a numerical implementation of the minimization process we show how we can obtain
density profiles in three dimensions for arbitraty potentials, and we test the limits of 
validity of the theory.
\end{abstract}

\maketitle
\section{Introduction}
The classical theory of fluids in thermal equilibrium is a highly developed discipline.
Along with specific physically motivated approximations have come tools of
more general validity and utility.
Studies of polymers have of course been extensive, and useful relationships between polymer melts
and fluids established \cite{1}, as well as between single polymers and fluids.
This paper represents an initial investigation, aimed at understanding which concepts borrowed from 
bulk fluid studies \cite{2,3} still remain relevant in a polymer setting.
Our focus will be on a single polymer chain confined by external forces, and to
minimize the needed information input, on single homopolymers.
We will also attend in the main to idealized models in which the polymer is simply a chain of unit monomers 
of a few degrees of freedom, but will indicate how this restriction can be rewardingly removed, 
on the way to a realistic polymer representation.

We will aim at both analytic simplicity and reasonable suitability for the
ultimately necessary computational procedures.
For the former, we will work in a ``monomer grand-ensemble''\cite{4,5} in which the number of monomers per 
polymer is distributed, but will show that this need not be a drawback.
For the latter, we will favor minimum principles to be able to better control computations.
This does restrict the category of systems to be studied e.g. to purely attractive pair interactions, 
and will be generalized in later work, soon to be reported, as such restrictions are removed.

\section{The Reference System}
\subsection{Notation}
Let us be a bit more explicit.
We have in mind an ordered chain of $N$ equivalent monomers, the $j^{th}$ being specified by its degrees 
of freedom $r_j$.
The order is maintained by a symmetric next neighbor interaction potential of Boltzmann factor $w(r_i,r_{i+1})$,
depending of course on the inverse temperature $\b$.
Any two monomers can also interact via an interaction Boltzmann factor $e(r_i,r_j) = \exp(-\b \f(r_i,r_j))$ and, 
crucially for the applications we have in mind, the polymer is constrained by an external potential $u(r_i)$.

It is now convenient to imagine that the homopolymer in question is both in thermal equilibrium and 
in number equilibrium, i.e. that it is the result of monomer addition and absorption from a bath 
of non-interacting monomers.
The reaction equilibrium is analogous to that of the grand canonical ensemble for a fluid, 
but with a significant difference.
Suppose that $Q^{(m)}$ is the monomer canonical partition function in its center of mass coordinate system,
$Q_N^{(p)}$ of the $N$-monomer polymer ($N\geq 1$, to recognize an object as a polymer).
Then if the full system contains $N'$ monomers, $N$ of which are bound together, in a volume $V$, the full
system partition function will be:
\beq 
Q^{tot}_{N'} = \sum_{N\geq 1} \frac{(V Q^{(m)})^{N'-N}}{(N'-N)!} Q^{(p)}_N.
\eeq
Since the monomers in the absence of a polymer would have a partition function:
\beq \label{Qm}
Q^{(m)}_{N'} = \frac{(V Q^{(m)})^{N'}}{N'!}
\eeq
that attributed to the polymer will take on the form:
\beqa 
\Xi^{(p)}[\z_0] &=& \lim_{N'\to\inf, N'/V \mathrm{fixed}} \frac{N'!}{(V Q^{(m)})^{N'}} Q^{tot}_{N'} \\
&=& \sum_{N\geq 1} \z_0^N Q_N^{(p)}, \qquad \mathrm{where} \;\;\z_0 = (N'/V)/Q^{(m)}. \label{partition_P}
\eeqa
The obvious analogy with a fluid grand partition function (with the weight $1/N!$ excised) will be very useful indeed.

The computation of $\Xi^{(p)}(\z_0)$, which generates all thermodynamics and expectations in a thermal ensemble,
is of course too general to be explicitely solvable, except in very special circumstances.
Let us therefore start with the evaluation of  $\Xi^{(p)}(\z_0)$ for what may be regarded as the backbone of the polymer,
that in which the arbitrary pair mutual interaction $\phi(r_i,r_j)$ is set equal to zero.
In this case, we have at once:
\beqa \label{rhoQ}
\z_0^N Q_N^{(p)} &=& \int \z(r_1) w(r_1,r_2) \z(r_2) \ldots \z(r_{N-1}) w(r_{N-1},r_N) \z(r_N) dr^N\\
\mathrm{where} && \z(r) = \z_0 e^{-\b u(r)} = e^{\b\mu(r)} \nonumber
\eeqa
or regarding $w(r,r')$ as the kernel of an integral operator w, $\z(r) \delta(r-r')$ as that of the 
diagonal operator $\z$, and the symbol $|1\>$ denoting the vector whose components are $1$,
\beq \label{partition_ref}
\Xi^{(p)}[\z] = \<1|(\z^{-1}-w)^{-1}|1\>,
\eeq
subject of course to convergence of the series (\ref{partition_P}).
The corresponding ``grand potential'' is:
\beq
\Omega^{(p)}[\z] = -\frac{1}{\b} \ln \<1|(\z^{-1}-w)^{-1}|1\>
\eeq
and as an immediate consequence the monomer density is given by
\beqa
n(r) &=& -\b \z(r) \frac{\d}{\d \z(r)}\Omega^{(p)}[\z] \\
&=& \<1|(\z^{-1}-w)^{-1}|r\>\<r|(\z^{-1}-w)^{-1}|1\>\z^{-1}(r)/\<1|(\z^{-1}-w)^{-1}|1\>  \label{9}
\eeqa

\subsection{Number Distribution}
The absence of the statistical  weight $1/N!$ suggests an unusual distribution of monomer number.
To verify this, we first observe from (\ref{9}) that if $\hat{N}$ denotes the monomer number in a 
given configuration, 
than 
\beqa
N=\<\hat{N}\> &=& \int n(r) dr = \<1|(\z^{-1}-w)^{-1}\z^{-1}(\z^{-1}-w)^{-1}|1\> / \Xi^{(p)} \\
&=& \<\z^{1/2}|(I - \z^{1/2} w \z^{1/2})^{-2}|\z^{1/2}\>/\<\z^{1/2}|(I - \z^{1/2} w \z^{1/2})^{-1}|\z^{1/2}\> \\
&=& 1/(1-\l_0) + \mathcal{O}\left(1/(1-\l_1)\right)
\eeqa
where $\l_0 = \l_{max} (\z^{1/2} w \z^{1/2})$ and $\l_1$ is the next largest
eigenvalue; $\l_1<\l_0$ in a confined system.
If $N$ is large, then
\beq
\Xi^{(p)}(\z) = \<\z^{1/2}|(I - \z^{1/2} w \z^{1/2})^{-1}|\z^{1/2}\>
\eeq
is dominated by the ``resonance'' at $\l_0$, so that if
\beq \label{eigeq}
(\z^{1/2} w \z^{1/2})\psi_{\l_0} = \l_0\psi_{\l_0}
\eeq
with normalized $\psi$, then
\beq
\Xi^{(p)}[\z] \sim \frac{1}{1-\l_0} \<\psi_{\l_0}|\z^{1/2}|1\>^2
\eeq

There are now two consequences.
On the one hand, we have 
\beqa
\<e^{i\theta \hat{N}}\> &=& e^{i\theta \z_0 \partial/\partial\z_0}\Xi^{(p)}[\z]/\Xi^{(p)}[\z] \\
&=& \Xi^{(p)}[\z e^{i\theta}]/\Xi^{(p)}[\z] \\
&\cong& (1-\l_0)/(1-\l_0 e^{i\theta})
\eeqa
so that the distribution of $\hat{N}$ is given by 
\beq
f(N) = \mathrm{coef}\, e^{i N \theta} (1-\l_0)/(1-\l_0 e^{i\theta}) = \l_0^N(1-\l_0)
\eeq
the very broad geometric distribution.
It would appear that the $N$-ensemble must give a very poor representation of a given $N$.
But on the other hand, we have from (\ref{rhoQ})
\beqa
Q_N &=& \z_0^{-N} \<\z^{1/2}|(\z^{1/2} w \z^{1/2})^{N-1}|\z^{1/2}\> \\
&\cong& \z_0^{-N} \l_0^{N-1}\<\psi_{\l_0}|\z^{1/2}\>^2,
\eeqa
so that if $F_N$ is the canonical Helmholtz free energy,
\beqa
\b F_N - \b \Omega^{(p)} &=& -\ln Q_N + \ln\Xi^{(p)}[\z] \\
&\cong& N \ln \z_0 - (N-1) \ln \l_0 - \ln(1-\l_0).
\eeqa
It follows that for any ($N$-independent) parameter variation
\beq \label{24}
\b(\d F_N - \d\Omega^{(p)}) = \left(\frac{1}{1-\l_0} - \frac{N-1}{\l_0}\right)\d\l
\eeq
which vanishes if $\z_0$ in (\ref{eigeq}) is chosen so that
\beq \label{25}
N = 1/(1-\l_0)
\eeq
We conclude that for this choice of $\l_0$, expectations at fixed $N$ and in the monomer number 
ensemble are in fact identical to leading order.

\subsection{Minimum Principle}
With the confidence that the large spread of monomer number does not detract from the 
usefulness of $\Omega^{(p)}$, we proceed.
The original form (\ref{partition_ref}) is simplest to use for the construction of an associated minimum 
principle.
It depends upon the fact if $K$ is positive semi-definite, and $a$ and $\psi$ (not to be confused
with $\psi_{\l}$, which we will not need again until later) are arbitrary,
then according to the Schwartz inequality, we have
\beqa 
\<a|K^{-1}|a\>\<\psi|K|\psi\> &=& \<K^{-1/2}a |K^{-1/2} a\>\<K^{1/2}\psi|K^{1/2}\psi\> \\
&\geq& \<K^{-1/2}a|K^{1/2}\psi\>^2 = \<a|\psi\>^2,
\eeqa
so that $\<a|K^{-1}|a\> \geq \<a|\psi\>^2/<\psi|K|\psi\>$.
Then indeed
\beq
\mathrm{Max}_{\psi} \frac{\<a|\psi\>^2}{\<\psi|K|\psi\>} = \<a|K^{-1}|a\> 
\qquad \mathrm{at} K\psi = \frac{\<\psi|K|\psi\>}{\<a|\psi\>}a,
\eeq
and in the context of (\ref{partition_ref}), we conclude that 
\beq \label{minimum_pr}
\beta \Omega^{(p)}[\z] = Min_{\psi}\left[\ln\<\psi|\z^{-1} - w|\psi\> - 2\ln \<1|\psi\>\right].
\eeq

\section{The Mean Field Strategy}
Our task is now to take into account the mutual interaction $\phi_1(r_i,r_j)$ which while 
typically fairly short range in space, can be very long range along the polymer chain.
We will confine our attention to purely attractive (negative definite) potentials, and to 
avoid confusion will set
\beq
\phi_1(r_i,r_j) = -\phi(r_i,r_j)
\eeq
where $\f$ is positive definite (as a continuous matrix).
The assumption of pure attraction distorts local properties of the polymer, but
permits large scale effects such as globularity to proceed unhindered.

At first, we need not restrict our attention to polymer chains.
Quite generally, if $W_0^{(N)}(r^N,\mu)$ is the suitably weighted Boltzmann factor
for the $N$-unit configuration in the absence of $\f$, but with local chemical potential 
$\mu(r)$, the partition function for the interacting system can be written as
\beqa \label{partition_int}
\Xi[\mu,\f] &=& \sum_{N}\int\ldots\int W_0^{(N)}(r^N,\mu) e^{\frac{\b}{2} \sum_{i,j}'\f(r_i,r_j)} dr^N \\
&=& \sum_N \int\ldots\int W_0^{(N)}(r^N,\mu) e^{-\frac{\b}{2}\sum_i\f_D(r_i)}
e^{\frac{\b}{2}\sum_{i,j}\f(r_i,r_j)} dr^N\\
&=& \sum_N \int\ldots\int W_0^{(N)}(r^N,\mu) e^{-\frac{\b}{2}\sum_i\f_D(r_i)}
e^{\frac{\b}{2}\int\int \hat{n}(r) \f(r,r')\hat{n}(r') dr dr'} dr^N
\eeqa
where $\sum'$ omits the $i=j$ contribution, $\f_D$ is the diagonal part of the matrix 
$\f(r_i,r_j)$, and 
\beq
\hat{n}(r) = \sum_i\delta(r-r_i)
\eeq
is the ``microscopic'' density.
The device of Kac, Siegert, Hubbard and Stratonovich \cite{6,7,8,9} is to represent the Gaussian in
(\ref{partition_int}) (in obvious notation) as a functional Laplace transform
\beqa
e^{\frac{\b}{2} \hat{n}\cdot\f\hat{n}} &=& \int e^{-\frac{\b}{2}v\cdot\f^{-1}v}
e^{-\b v\cdot \hat{n}} Dv \;/ \int e^{-\frac{\b}{2}v\cdot\f^{-1}v} Dv \\
&=& \int e^{-\b \sum_i v(r_i)} e^{-\frac{\b}{2} v\cdot \f^{-1} v}Dv \;/
\int e^{-\frac{\b}{2}v\cdot\f^{-1}v} Dv
\eeqa
Since $\sum_N\int\ldots\int W_0^{(N)}(r^N,\mu) dr^N = \Xi_0[\mu]$, eq.(\ref{partition_int}) can 
thereby be rewritten as
\beq \label{partition_v}
\Xi[\mu,\f] = \int\Xi_0[\mu - \frac{1}{2}\f_D-v] e^{-\frac{\b}{2} v\cdot\f^{-1}v}Dv \;/
\int e^{-\frac{\b}{2}v\cdot\f^{-1}v} Dv
\eeq
with a possible interpretation that the interaction $-\f$ has been replaced by an ensemble average
over a fluctuating external field $v(r)$, serving as a sort of graviton shutting back and forth 
between units.

The kernel of (\ref{partition_v}) is a Boltzmann factor in field space, and so we may define a field 
average as
\beq \label{gaussian_av}
\<\< G[v] \>\> = \int G[v] \Xi_0[\mu - \frac{1}{2}\f_D-v] e^{-\frac{\b}{2} v\cdot\f^{-1} v}Dv \;/\Xi[\mu,\f]
\eeq
A very suggestive consequence of this notation follows from the observation that for the density $n(r)$,
\beqa \label{partial_n}
n(r) &=& \frac{1}{\Xi[\mu,\f]}\frac{\d}{\d \b\mu(r)} \Xi[\mu,\f] \\
&=& -\int \frac{\d}{\d \b \mu(r)} \Xi_0[\mu - \frac{1}{2}\f_D-v]
e^{-\frac{\b}{2} v\cdot\f^{-1} v}Dv \;/\Xi[\mu,\f] \int e^{-\frac{\b}{2}v\cdot\f^{-1}v} Dv,
\eeqa 
or integrating by parts in v-space (assuming the absence of boundary terms)
\beqa
n(r) &=& - \int \f^{-1}v(r) \Xi_0[\mu - \frac{1}{2}\f_D-v] e^{-\frac{\b}{2} v\cdot\f^{-1} v}Dv \;/
\Xi[\mu,\f] \int e^{-\frac{\b}{2}v\cdot\f^{-1}v} Dv \\
&=& -\f^{-1} \<\< v(r) \>\>,   \label{density_v}
\eeqa  
identifying $-\f n(r)$ as the ``mean field'' $\<\< v(r) \>\>$.
It is the approximate computation of this mean field that we must attend to, directly or indirectly.

If the field Boltzmann factor of (\ref{gaussian_av}) is sharply peaked about a function $\bar{v}(r)$, 
than of course (\ref{density_v}) becomes simply
\beq \label{density_phi}
n(r) = -\f^{-1}\bar{v}(r),
\eeq 
and the field $\bar{v}$ must now satisfy ($\Omega$ is again the grand potential $-\frac{1}{\b}\ln\Xi$)
\beq  \label{partial_O}
\frac{\d}{\d \bar{v}(r)}\left(\Omega_0[\mu -\frac{1}{2}\f_D-\bar{v}]\right) 
+ \frac{1}{2}\bar{v}\cdot\f^{-1}\bar{v} = 0,
\eeq
or simply
\beq
n_0(r|\mu -\frac{1}{2}\f_D-\bar{v}) = -\f^{-1}\bar{v}(r),
\eeq
the unsurprising result (comparing with (\ref{density_phi})) that $n(r)$ is equal to the ``bare'' 
density $n_0$ in the presence of $\bar{v}$, to within an additional shift of $\frac{1}{2}\f_D$.
But in addition, we now have in leading order approximation the very explicit
\beq \label{O1}
\Omega_1(\mu,\f) = \Omega_0[\mu -\frac{1}{2}\f_D-\bar{v}] + \frac{1}{2}\bar{v}\cdot\f^{-1}\bar{v}
\eeq
which, since (\ref{partial_O}) represents the minimization of the kernel of (\ref{partial_n}), 
takes the variational form
\beq \label{minimum}
\Omega_1(\mu,\f) = \mathrm{Min}_v \Omega_0[\mu -\frac{1}{2}\f_D-v] + \frac{1}{2}v\cdot\f^{-1}v.
\eeq
Furthermore, consistency is established by noting that by virtue of (\ref{minimum}), $\Omega_1$
of (\ref{O1}) implies
\beq
n(r) = -\frac{\d \Omega_1}{\d\mu(r)} = -\f^{-1}\bar{v}(r),
\eeq
reproducing (\ref{density_phi}).

To apply (\ref{minimum}) to the single homopolymer under discussion, we need
only insert (\ref{minimum_pr}), 
serving as $\Omega_0$, into (\ref{minimum}).
We then have
\beq \label{minimum_v_psi}
\b \Omega_1[\mu,\f] = \mathrm{Min}_{v,\psi}\left[\frac{\b}{2} v\cdot\f^{-1} v - 2 \ln\<1|\psi\>
+ \ln \<\psi|e^{-\b(\mu'-v)}-w|\psi\>\right]
\eeq
where $\mu'(r) = \mu(r) - \frac{1}{2} \f_D(r)$.
Eq. (\ref{minimum_v_psi}) can be simplified by finding and eliminating the field $\psi$, and replacing 
$v$ by $-\f n$.
To do so we observe that
\beq \label{partial_e_zero}
0 = \frac{\d\Omega_1}{\d v} = \f^{-1} v + e^{-\b(\mu'-v)}\psi^2 \;/
\<\psi|e^{-\b(\mu'-v)}-w |\psi\>
\eeq
Integrating over the implicit r, with $N = \int n(r) dr$, then
\beq
N = \frac{\<\psi|e^{-\b(\mu'-v)}|\psi\>}{\<\psi|e^{-\b(\mu'-v)}-w |\psi\>},
\eeq
reducing (\ref{partial_e_zero}) to 
\beq \label{implicit_density}
n = N e^{-\b(\mu'-v)} \psi^2 \;/\<\psi|e^{-\b(\mu'-v)}|\psi\>
\eeq
Since (\ref{implicit_density}) is homogeneous of degree $0$ in $\psi$, we are free to adopt the
normalization
\beq \label{normalization}
\<\psi|e^{-\b(\mu'-v)}|\psi\> = N
\eeq
converting (\ref{implicit_density}) to
\beq \label{54}
n = e^{-\b(\mu'-v)}\psi^2
\eeq
(leading back to (\ref{normalization})), and consequently replace (\ref{minimum_v_psi})
by the simple 
\beq \label{O1_min}
\b \Omega_1[\mu,\f] = \mathrm{Min}_n\left[\frac{\b}{2} n\cdot\f n -2 \ln\<1|n^{1/2} 
e^{\frac{\b}{2}(\mu'+\f\cdot n)}\> + \ln\<n^{1/2}|\left(I - e^{\frac{\b}{2}(\mu'+\f\cdot n)}
w e^{\frac{\b}{2}(\mu'+\f\cdot n)} \right)| n^{1/2}\>\right].
\eeq
appearing as a $\mu$-dependent density functional, reminiscent of
``statistical models'' \cite{10,11,12}, of polymers.
Eq.(\ref{O1_min}) is our main result, valid in the mean field level.
The role of the interaction $-\f$ is transparent: the external potential is 
augmented by the mean field :$\mu \to \mu'+\f n$, and the explicit energetic component 
$-\frac{1}{2} n\cdot\f n$ subtracted out.
Eq.(\ref{O1_min}) is of course a density functional representation with the density profile 
dependence on $\mu$ determined by dropping the implicit $\f$ and making $\mu$ explicit, by
\beq
\d\Omega_1[\mu, n] \;/\d n(r) = 0.
\eeq
It is an extension of the more common form in which one is given a free energy $F[n]$ in 
terms of which $\mu = \d F(n)/\d n$, from which it follows that
\beq
\d \Omega(\mu,n) \;/\d n = 0,
\eeq
where $\Omega[\mu,n] = F[n] - \mu\cdot n$.
A corresponding format is obtainable here as well, but at the cost of increased complexity.

\section{Numerical Implementation}
To exemplify the usefulness of the mean field minimization strategy, we have
developed a numerical implementation of the minimization principle
(\ref{O1_min}).

From preliminary simulation studies we had noticed how in the absence of a
repulsive core interaction between monomers we could not converge to a stationary density.
Instead the density, and therefore the total particle number, either kept on growing 
or shrank to zero, depending on the value of the external chemical potential.
This numerical instability occures as a result of the closeness to the resonance.
In these simulations we noticed, on the other hand, that despite the fact that
the density itself kept changing, the global shape of density profile
reached a steady state.
We therefore decided to study the normalized density $p(r) = n(r)/N$.
Once we had formulated the problem consistently in terms of $p$ all
the numerical instabilites desappeared and we were able to determine the corresponding 
stady state densities and total particle numbers.

In terms of $p$ the minimization functional (\ref{O1_min}) becomes
\beq
\Omega_1[\mu,\f] = \mathrm{Min}_p\left[\frac{N^2}{2}p\,\cdot\phi p - 2 \ln\<1|p^{1/2}
e^{\frac{\b}{2}(\mu'+N\phi\psi)}\>+\ln\<p^{1/2}|\left(I-e^{\frac{\b}{2}(\mu'+N\phi\psi)}w 
e^{\frac{\b}{2}(\mu'+N\phi\psi)}\right)|p^{1/2}\>\right].
\eeq
Once we have determined $p(r)$, the expression to compute the particle number
and the true density are
\beqa 
n(r) &=& \frac{p(r)}{\<p(r)^{1/2}[I - \z w \z]p(r')^{1/2}\>} \\
N &=& \<p(r)^{1/2}[I - \z w \z]p(r')^{1/2}\>^{-1}
\eeqa

The minimization is done using a Metropolis Monte Carlo algorithm at fixed
temperature, starting from a suitable ansatz for the functional form of $p(r)$.
We can freely impose the form of the next neighbor potential, and therefore
$w(r,r')$, the external confining potential $u(r)$, and the long range 
interaction $\phi(r,r')$.
The main numerical difficulty arises from the need of evaluating a functional
involving 6-dimensional integrals nesting 3-dimensional integrals, such as
\beq
\int \sqrt{p({\bf r})}e^{\frac{\b}{2}\left(\mu'({\bf r})+N\int\phi({\bf r},{\bf
    r''})p({\bf r''})d{\bf r''}\right)}w({\bf r},{\bf r'})e^{\frac{\b}{2}\left(\mu'({\bf r'})+N\int\phi({\bf r'},{\bf
    r''})p({\bf r''})d{\bf r''}\right)} \sqrt{p({\bf r'})}d{\bf r}\,d{\bf r'}
\eeq 
It turns out it is prohibitive to try to evaluate these terms directly by the
use of a simple grid. Already when using 10 discretization points simulations 
are too slow, while the precision is very poor.
To achieve higher accuracy in the integration, we use a Legendre-Gauss quadrature
method with either 6 or 8 points in each linear direction.

As a test system we have chosen a square well external potential with $\mu =
-\mu_0$ inside; a harmonic n.n. potential, leading to $w({\bf r},{\bf r'})=
\exp(-C_h (|{\bf r} - {\bf r'}|-\rho_1)^2)$; a long-range
step potential $\phi=C_{lr} \Theta(x-\rho_2)$, where $\mu_0$, $C_h$, $C_{lr}$
are energy constants, and $\rho_1$ and $\rho_2$ are characteristic lengths.
In the following we take the system size to be $L=1$, $C_h=5$, and $\rho_1=0.2$.
As ansatz for the probability density we take the composition of three
Gaussian in the three spatial directions, for a total of 6 parameters:
$B_x$,$B_y$,$B_z$,$X_0$,$Y_0$ and $Z_0$.
\beq
p(x,y,z) = A\; e^{-B_x(x-X_0)^2}e^{-B_y(y-Y_0)^2}e^{-B_z(z-Z_0)^2}
\eeq
At each iteration of the algorithm the normalization of $n$ is verified so
that $\int p(x,y,z) dx\, dy\, dz = 1$, and the value of A is changed
accordingly.

We first look at the reference system, i.e. zero long-range interaction, figure 1.
\begin{figure}[t]  
\begin{center}
\includegraphics[width=8cm]{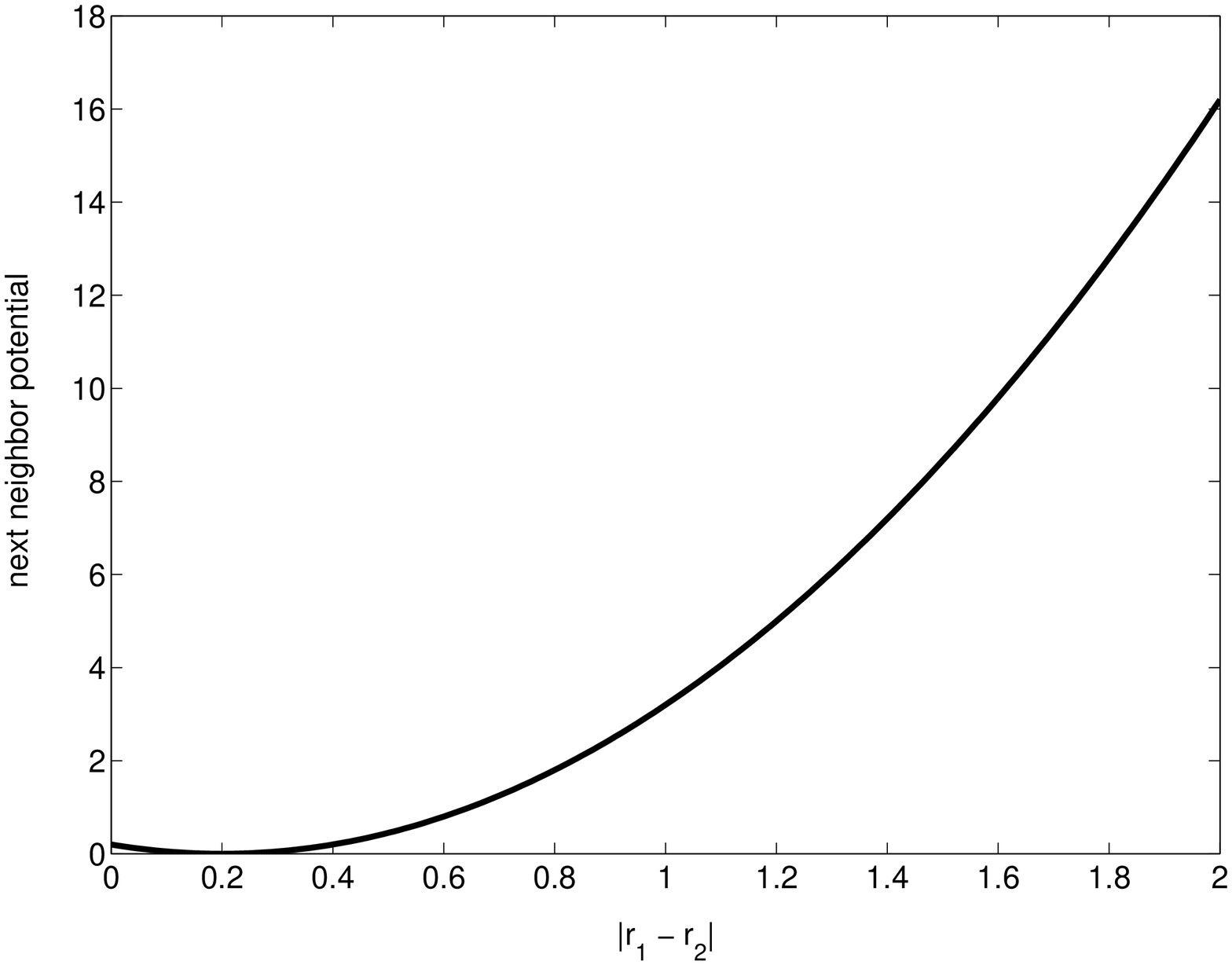} 
\includegraphics[width=6cm]{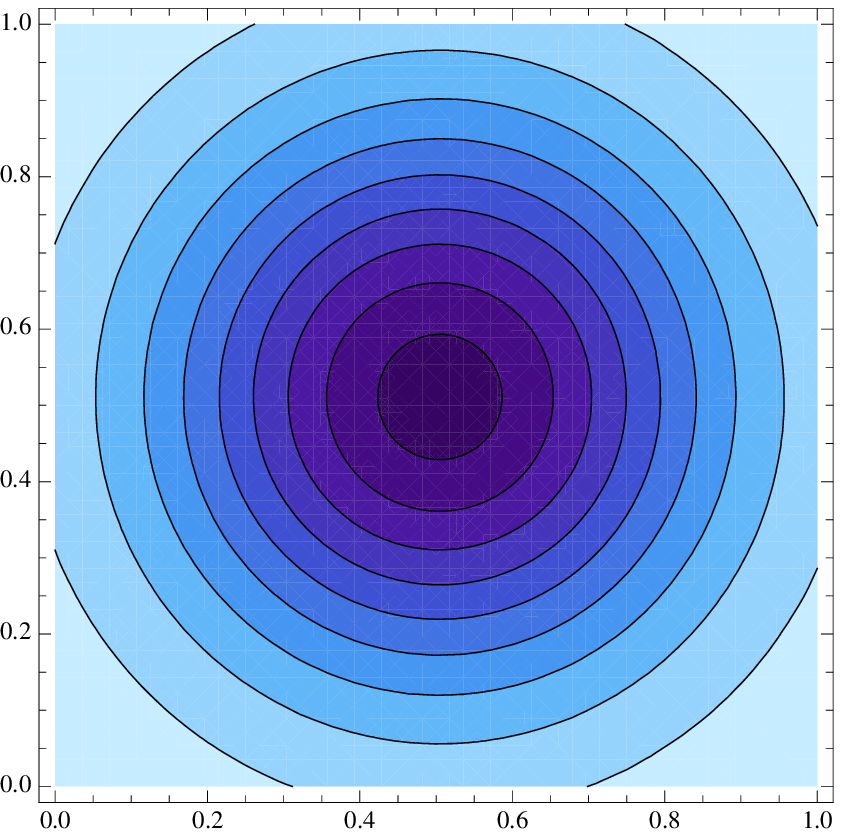} 
\caption{{\small Left: Interaction between neighboring particles along the chain:
    harmonic potential only. Right: Slice of the probability density profile taken at $z=L/2$.}} \label{fig_reference}
\end{center}
\end{figure}
As expected the probability density converges to a single Gaussian centered in
the middle of the box, with $X_0 \sim Y_0 \sim Z_0\sim L/2$ and $B_x \sim B_y \sim B_z$.
When the attractive potential $\phi$ is turned on the Gaussian becomes more
peaked (figure 2).          ).
\begin{figure}[t] \label{fig_phi}
\begin{center}
\includegraphics[width=8cm]{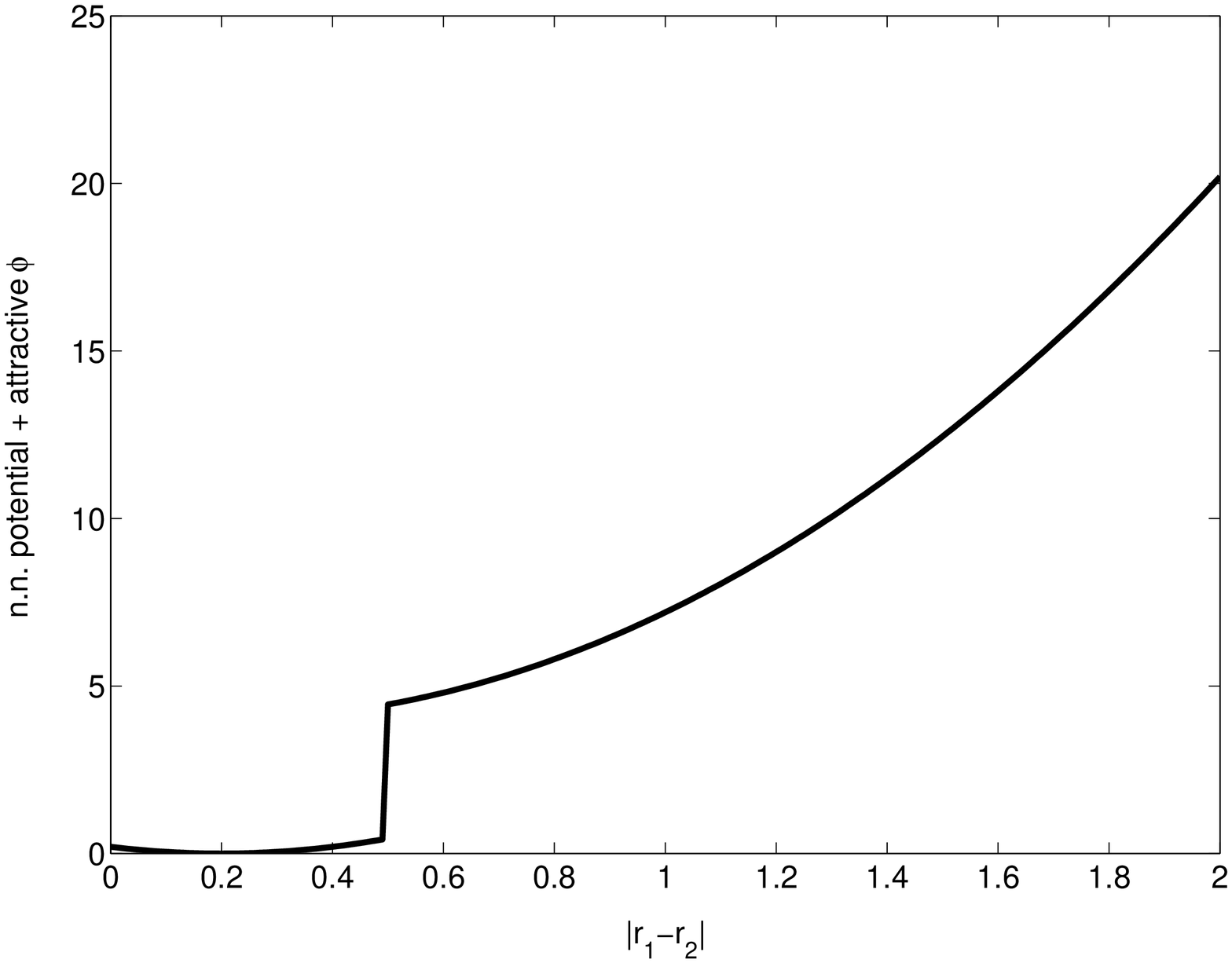} 
\includegraphics[width=6cm]{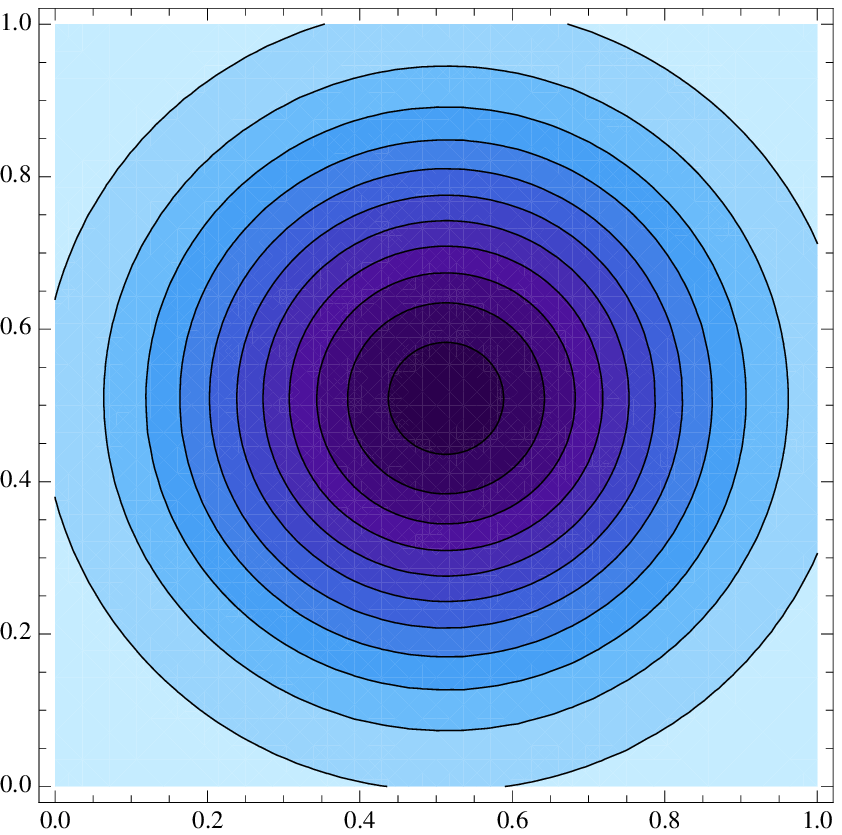} 
\caption{{\small Left: Interaction between neighboring particles along the chain:
    harmonic potential, attractive long-range interaction ($C_l=4$),
    $\rho_2=0.5$. Right: Slice of the probability density profile taken at $z=L/2$.}} 
\end{center}
\end{figure}

Next, we introduced an asymmetry in the external potential, like a
gravitational potential along $z$: $\mu = \mu_0 -g z$.
As expected the minimization of the functional converges to a probability
density off-centered with respect to $z$ and centered in the middle of the box
with respect to $x$ and $y$ (figure 3).
\begin{figure}[t]  \label{fig_gravity}
\begin{center}
\includegraphics[width=6cm]{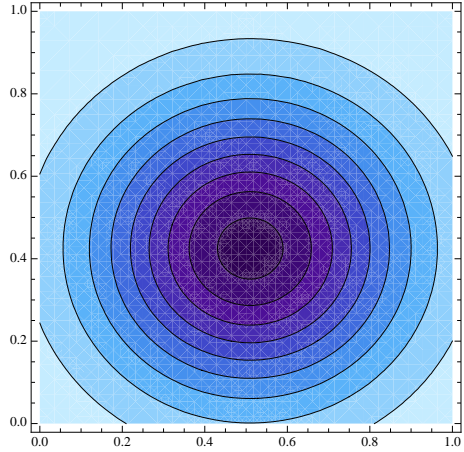} 
\caption{{\small Probability density profile at $z=L/2$ in the presence of an external
    graviational field ($g=1$, $C_l=2$)}} 
\end{center}
\end{figure}

Even if the theoretical derivation of eq.(\ref{O1_min}) is valid only for
strictly positive definite $\phi$, in the algorithm we can invert the sign of
the long range interaction and make it repulsive, as well as attractive.
We tried to go beyond the known validity of our approximation, taking a
repulsive long-range interaction.
Our method still holds, provided we consider repulsive interactions
which are not too strong.
As expected a repulsive $\phi$ gives a less peaked density compared to the
reference system (figure 4).
\begin{figure}[t]  
\begin{center}
\includegraphics[width=8cm]{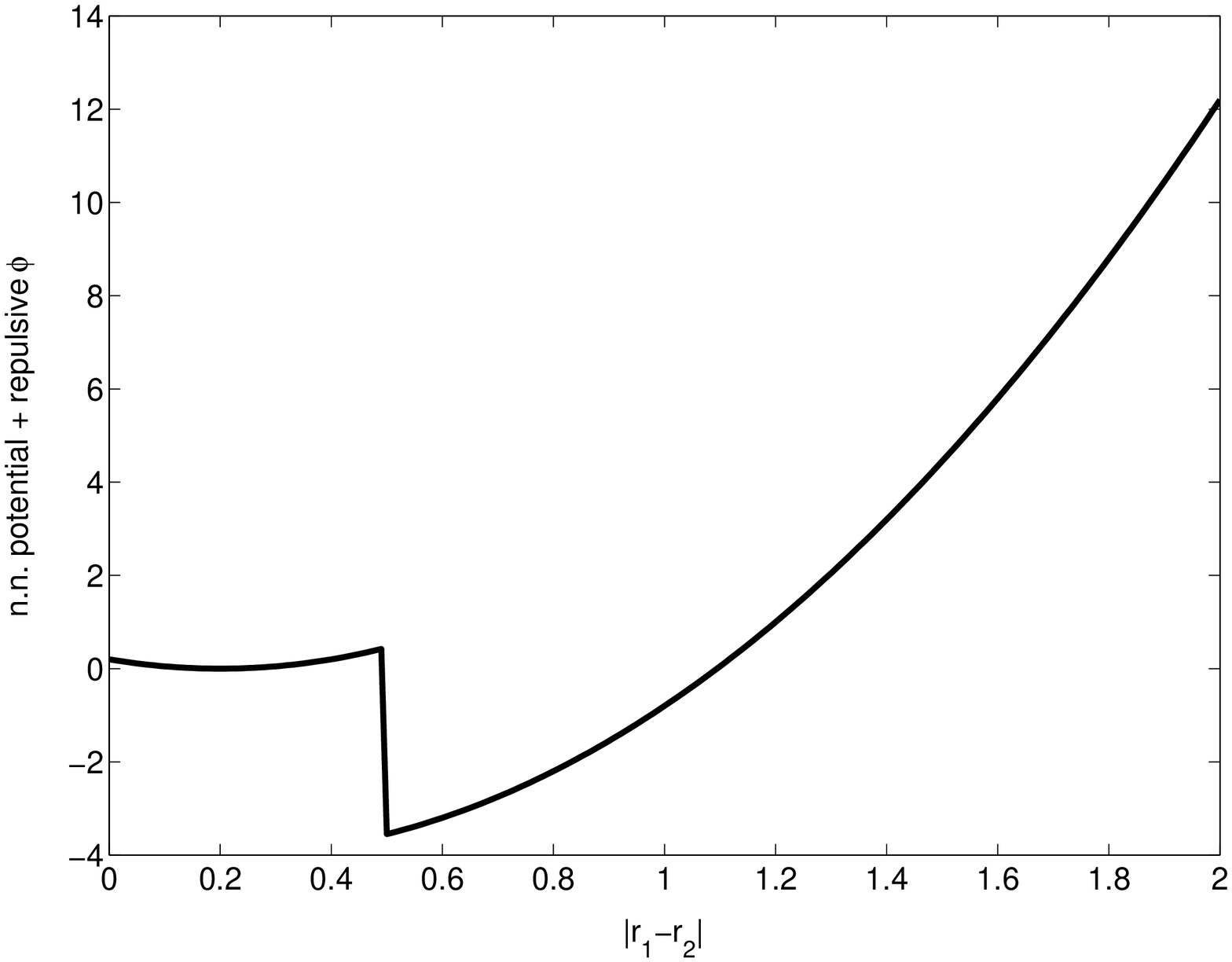}
\includegraphics[width=8cm]{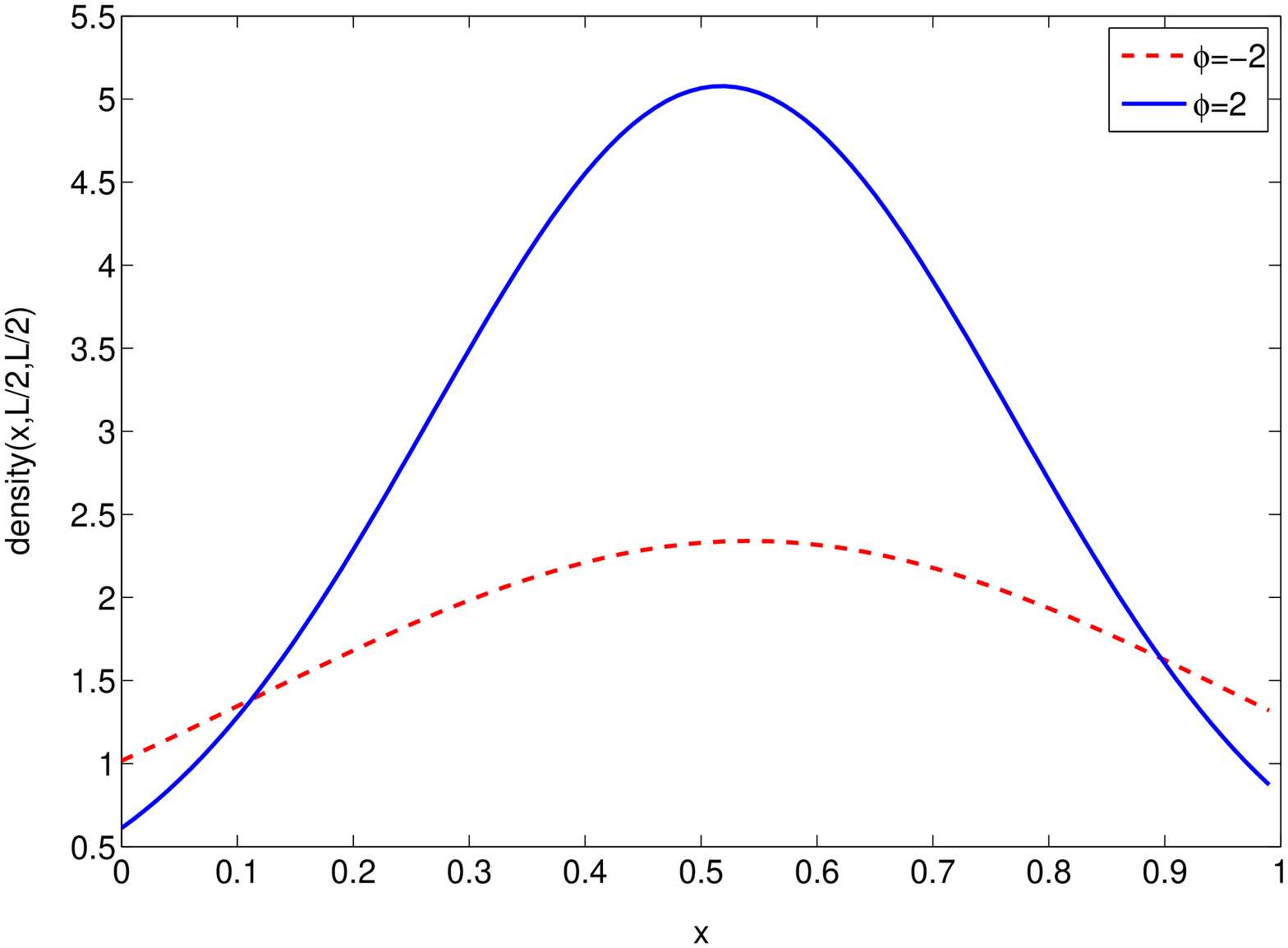}
\caption{{\small Left: Interaction between neighboring particles along the chain:
    harmonic potential and repulsive long-range interaction ($C_l=-2$). Right:
probability profiles at $y=z=L/2$ with (red) and without (blue) the repulsive core.}} \label{fig_3}
\end{center}
\end{figure}

\section{Limit of tight confinement}
We have seen, in Sec. II B, that in the absence of non-neighbor interactions,
the large $N$ monomer distribution is determined by the ``resonance state''
$\psi_{\l_0}$ satisfying
\beq \label{psi_0}
(\z^{1/2} w \z^{1/2})\psi_{\l_0}=\l_0\psi_{\l_0},
\eeq
where $\l_{MAX}=\l_0 \sim 1$.
Explicitely, since 
\beqa
\<r|\left(\z^{-1}-w\right)^{-1}|1\> &=& \<r|\z^{1/2}\left(I - \z^{1/2} w
\z^{1/2}\right)^{-1}\z^{1/2}|1\> \nonumber \\
&\sim &\frac{1}{1-\l_0}\<r|\z^{1/2}\psi_{\l_0}\>\<\psi_{\l_0}|\z^{1/2}|1\> 
\eeqa
we have under these circumstances, from (\ref{9})
\beq
n(r) \sim \frac{1}{1-\l_0}\psi_{\l_0}(r)^2,
\eeq
or since $\psi_{\l_0}(r)\sim \psi_1(r)$,
\beq \label{density_p_N}
n(r) \sim \frac{\psi(r)^2}{1-\l_0} = N \psi_1(r)^2
\eeq
In other words, increasing the chain length increases the mean density
uniformly, as if the floppy chain is simply winding around more under the same
confinement.
This uninteresting behavior is refined by two interactions that have been
ignored in getting (\ref{density_p_N}).
First is the stiffness of successive pair orientations, equivalent to the
monomers having coupled orientational degrees of freedom, a topic that has been
addressed to some extent in this format in the past and will be attended to
more forcefully in the future. 
Second is the effect of non-next neighbor interactions, which we have here
studied in a preliminary fashion.
The mean field $v(r)$ that has been enountered will also have the effect of
correlating pair orientations, and will of course alter the nature of the long
chain resonant state.
Let us see how this works.
To keep the extrapolation from (\ref{psi_0}-\ref{density_p_N}) transparent,
let us rewrite (\ref{minimum_v_psi}) (using $\z=e^{\b\mu'}$) as, 
\beqa \label{minimum_bar}
\Omega_1(\z,\phi) &=& \mathrm{Min}_{v,\psi}[\frac{1}{2}v\cdot \phi^{-1} v - 2
  \ln\<(\r e^{-\b v})^{1/2}|(\r^{-1}e^{\b v})^{1/2}\psi\> \nonumber \\
&+& \ln\<(\r^{-1}e^{\b
    v})^{1/2}\psi|I - (\r e^{-\b v})^{1/2} w (\r e^{-\b
    v})^{1/2}|(\r^{-1}e^{\b v})^{1/2}\psi\>]\nonumber \\
&=&\mathrm{Min}_{v,\bar{\psi}}\left[1\over 2 v\cdot\phi^{-1} -2 \ln\<(\r e^{-b
    v})^{1/2}|\bar{\psi}\> + \ln\<\bar{\psi}|I - (\r e^{-b v})^{1/2} w (\r
  e^{-\b v})^{1/2}|\bar{\psi}\>\right]
\eeqa
Hence, according to (\ref{normalization}), $\bar{\psi} = (\z^{-1} e^{\b v})^{1/2}\psi$ is
normalized, $\<\bar{\psi}|\bar{\psi}\>=1$ and from (\ref{54}),
\beq \label{N_psi_bar}
n = N \bar{\psi}^2
\eeq
Near resonance is now signaled by the approximate validity of
\beq \label{approximation}
\left(I -(\z e^{-\b v})^{1/2} w (\z e^{-\b v})^{1/2}\right)
\bar{\psi}=\bar{\lambda}\bar{\psi}
\eeq
with small $\bar{\lambda}>0$.
Eq. (50) then tells us at once that using the exact
\beq \label{neg_n}
\phi^{-1}v = -n,
\eeq
we have 
\beq \label{lambda_bar}
N = 1/\bar{\lambda}
\eeq

There are two ways of making use of the approximation \ref{approximation}.
Most directly, we substitute (\ref{N_psi_bar}-\ref{lambda_bar}) into (\ref{minimum_bar}), obtaining
\beq
\b\Omega_1\sim \mathrm{Min}_n\left[\frac{1}{2} n\cdot \phi n - 2 \ln\<(\z
  e^{\b\phi n})^{1/2}|n^{1/2}\>\right]
\eeq
This is obviously too sweeping an approximation: by using what is
effectively a first order correction in the argument of a variational
principle, the role of the next-neighbor $w$ in $\b\Omega_1$ has vanished. But
we can pick up the next order by working instead at the ``profile equation''
level.
It is only necessary to substitute (\ref{N_psi_bar}, \ref{neg_n}, \ref{lambda_bar}) 
directly into (\ref{approximation}) to rewrite the latter as
\beq
(\z e^{\b v})^{1/2}w(\z e^{-\b v})^{1/2} n^{1/2} = \left(1-\frac{1}{N}\right) n^{1/2},
\eeq
and hence as
\beqa \label{Lambda_n}
e^{-\frac{\b}{2}(u - \phi n)}w e^{\frac{\b}{2}(u-\phi n)}n^{1/2} &=&
\left(1-\frac{1}{N}
\right)e^{-\b\mu'_0} n^{1/2}\nonumber \\
&\equiv&\Lambda n^{1/2}
\eeqa
($\mu'_0$ is the global chemical potential).
The operator in (\ref{Lambda_n}) $(x^2)$ has a positive kernel, hence by Jentchke's
extension of Perron-Frobenius \cite{13}, the positive eigenfunction is unique (up to a multiplicative
constant) and $\Lambda$ is real and maximal.
This leads to what is in principle a simple numerical iteration: start
eg. with $\phi=0$, compute the eigenfunction $n^{1/2}$ and normalize it to get
$\int n(r) dr = N$.
At the $k^{th}$ stage of the M-fold iteration, replace $\phi$ by $(k/M)\phi$ and
repeat the process using the current function $n$. 
An alternative strategy is to parameterize $n$ and determine the parameters by
Galerkin, i.e. integrate (\ref{Lambda_n}) with weight function and solve the resulting
algebraic equation.

\section{Concluding remarks}
In this inital study, we have made one major approximation and several simplifying restrictions.
The approximation is of course that of mean field, or selective neglect of fluctuations.
On the assumption that fluctuations are Gaussian to leading order (examples in which this is not the 
case are far from rare, see e.g. \cite{A}), a correction sequence is in primciple routine: we expand
$\ln\Xi_0$ in (\ref{partition_v}) and (\ref{density_v}) around $\bar{v}$ of (\ref{partial_O}). 
Using $\d^2\Omega_0/\d v(r)\d v(r')= \d n_0(r)/\d v(r')|_{\bar{v}}$,
\beqa
&&\Xi_0[\mu -\frac{1}{2}\phi_D - v] \exp\{-\frac{1}{2} \b\, v\cdot \phi^{-1} v\} = \nonumber \\
&&= \exp\left\{-\b \left( \Omega_0[\mu -\frac{1}{2}\phi_D - \bar{v}] + \frac{1}{2}\bar{v}\cdot \phi^{-1} \bar{v}
+ \frac{1}{2}(v-\bar{v})\cdot\left[\frac{\d n_0}{\d v}|_{\bar{v}} + \phi^{-1}\right] (v-\bar{v}) + 
\ldots \right) \right\}
\eeqa
It readily follows (see e.g. \cite{B}) that the density profile is given to the next order by simply 
averaging the density over the Gaussian field fluctuation:
\beq 
n(r) = \frac{\int n_0[\mu -\frac{1}{2}\phi_D - \bar{v} - \Delta] 
e^{-\frac{1}{2}\left(\frac{\Delta^2}{M^{-1}(r,r)}\right)}d\Delta}
{\int e^{-\frac{1}{2}\left(\frac{\Delta^2}{M^{-1}(r,r)}\right)}d\Delta}
\eeq
where $M(r,r') = \phi^{-1}(r,r')-n_0'[\mu -\frac{1}{2}\phi_D - \bar{v} - \Delta]$;
here, $\Delta$ is the field amplitude fluctuation at $r$.
The simplifying assumptions attend more directly to the physics, and these assumptions depend
very much on the nature of the system to be studied.
Taking these assumptions in order, we first emebedded out system in a monomer grand ensemble.
Since preprocessing assures that in practice one does not deal with the resulting extreme polydispersity, a 
fixed $N$ ensemble is more relevant than fixed $\zeta$. 
The corresponding inverse mapping has been attended to on numerous occasions (see e.g. \cite{C}). 
The same formalism is indeed avaliable here (hinted at in eqs.(\ref{24}),(\ref{25})).

Another restriction was to attractive long-range forces. We found however that the profile equation
could indeed be pushed into the partially repulsive regime, although the validity of the minimum principle
was in question; this is closely related to the functional Fourier transform for the repulsive component,
likewise under unceirtain control.
An alternative approach lies in the use of the mean spherical model \cite{D,E} and its extensions.
This is the aim of ongoing research.

Of course, there is the implicit assumption that pair forces suffice, whereas the action of pairs on 
singlets is a frequent important observation, leading e.g. to dihedral angle dependence in chains.
Typically (see e.g. \cite{F} for a very primitive example) one can simply create a multi-unit monomer to 
encompass only such forces, which then appear once more as pair forces.

Most blatantly, our approach has been restricted to homopolymers. Since the set of degrees 
of freedom of a monomer can include monomer type, this is no restriction at all if one is 
studying the effect on the full 
population of an assumed relative frequency of occurrence of next neightbors hetero-pairs.
However, if we attribute a sense to the chain and the AB frequency differs from the BA frequency, the very
convenient symmetry of the operator $w$ is lost, and with it, the possibility e.g. of a specific long
sequence of monomer types. 
The case of non-symmetric $w$ has indeed been studied \cite{G}, but exercising the kind of control that we have 
in our current formulation remains a challenge.

\end{document}